\documentclass{appolb}
\usepackage{epsfig}

\begin{document}
\title{Heavy flavour and quarkonium at forward and central rapidity in p-p collisions with ALICE at LHC%
\thanks{Presented at Epiphany 2011}%
}
\author{R. Tieulent on behalf of the ALICE collaboration
\address{Universit\'e de Lyon, Universit\'e Lyon 1, CNRS/IN2P3, Institut de Physique Nucl\'eaire de Lyon, Villeurbanne, France}
}
\maketitle
\begin{abstract}

Quarkonia and open heavy flavour production are crucial to study the properties of the nuclear matter at high energy densities and of the Quark Gluon Plasma (QGP). In proton-proton collisions at LHC, the measurement of their production cross-section will allow to test perturbative QCD calculations in a new energy regime and provide the baseline for measurements in heavy ion collisions. In the latter, as they experience the full evolution of the collision, heavy quarks are ideal probes to study the property of the system. 
During the 2010 campaign of the LHC, ALICE took proton-proton data at $\sqrt{s}$=7~TeV and Pb-Pb data at $\sqrt{s_{nn}}$=2.76~TeV. The first physics results on heavy flavors production in p-p collisions are presented combining hadronic and (semi-)leptonic channels, both at central and forward rapidity.

\end{abstract}
  
\section{Introduction}

ALICE (A Large Ion Collider Experiment) \cite{ALICEatLHC} is the experiment dedicated to study  the properties of strongly interacting matter at extreme energy densities ($>$~10~GeV/fm$^3$), as those which are expected to be reached in Pb-Pb collisions at the LHC. Lattice QCD calculations predict that under this condition a deconfined Quark Gluon Plasma could be formed. In the new energy domain reached by the LHC, heavy quarks are abundantly produced. The study of their production, propagation and hadronization in the hot and dense nuclear matter created in heavy ion collisions can provide valuable information about the properties of the system. In fact, heavy quarks are produced in the first stage of the collisions and then they coexist with the surrounding medium due to their long lifetime. For example, the energy density of the created medium can be inferred by an energy loss measurement of heavy quark. The study of the production of heavy quarkonia, $J/\psi$, $\psi'$ ($c\bar{c}$ states), $\Upsilon$, $\Upsilon'$ and $\Upsilon''$, ($b\bar{b}$ states) is particularly interesting, because it can lead to a measurement of the temperature of the system. It was originally proposed that quarkonium resonances will dissociate by color screening in the QGP \cite{matsui}, thus a suppression of quarkonium production in nucleus-nucleus collisions compared to proton-proton collisions was predicted as a signature of the QGP formation. However, it is now recognized that in order to interpret $J/\psi$ production as a QGP probe one has to consider cold nuclear matter effects such as initial state energy loss and shadowing, as well as charm quark energy loss, co-mover interactions, corrections for feed-down from higher mass charmonium states, and secondary production mechanisms, such as recombination of initially uncorrelated $c\bar{c}$ pairs \cite{Rapp}.
In this contribution we present results based on the data collected by the ALICE experiment in proton-proton collisions at $\sqrt{s}$=7 TeV. 

\section{Heavy flavour detection by the ALICE experiment}

The detailed description of the ALICE experimental setup, which was designed to deal with charged particle multiplicity of few thousand particles per unit of rapidity, can be found in reference \cite{ALICEatLHC}. 

In the central region ($|\eta|<$0.9), the heavy flavour capability of ALICE relies in a high granularity tracking system made of the Inner Tracking System (ITS), the Time Projection Chamber (TPC) and the Transition Radiation Detector (TRD).  The particle identification is performed via dE/dx measurement in the TPC, via time of flight measurement in the Time Of Flight detector (TOF), and via electron/pion discrimination operated by the TRD\footnote{In the presented analyses the TRD is not used in the PID.}. These central detectors are embedded in a solenoidal magnetic field of 0.5 T.
Open charm is studied via the reconstruction of hadronic decays. In the following we will present the results concerning the channels  D$^0\rightarrow K^- \pi^+$ and D$^+\rightarrow K^- \pi^+\pi^+$. The open charm and open beauty are also studied via their inclusive semi-electronic decay channels D,B$\rightarrow e + X$. Finally, the quarkonia can be detected via their di-electronic decay channel. In this report we present the inclusive J/$\psi$ cross section.

In the forward region, the ALICE muon spectrometer covers the rapidity range -4.0$<\eta<$-2.5. The muon spectrometer consists of an absorber to filter the hadronic background, a tracking system made of ten cathode pad chambers, and four planes of trigger chambers made of Resistive Plate Chambers (RPC) placed downstream of an iron wall of 1.2 m, which acts as a muon filter. The spectrometer makes possible the study of the open charm and beauty in the inclusive semi-muonic decay channel (D, B$\rightarrow \mu + X$). In this report we present the p$_t$-differential cross section of inclusive muons from heavy flavour decays in 2$<p_{\rm T}<$6.5~GeV/c. Quarkonia states are measured via their di-muonic decay channel. In the following we present the p$_t$-differential cross section of inclusive J/$\psi$.

\section{Data sample}

At central rapidity, the p-p data at $\sqrt{s}$=7~TeV were collected with a
minimum bias (MB) trigger requiring at least one hit in the SPD (Silicon Pixel Detector, forming the two innermost layers of the ITS) or a signal in at least one of the two VZERO detectors (two scintillator hodoscopes placed on either side of the interaction region at z=3.3 m
and z=-0.9 m, covering the pseudorapidity regions 2.8 $<\eta<$5.1 and -3.7$<\eta<$-1.7). This trigger corresponds approximately to at least one charged particle over 8 units of pseudorapidity. The events were in coincidence with signals from two beam pick-up counters, one on each side of the interaction region, signing the passage of proton bunches. The absolute normalization has been done using minimum-bias cross section extracted from the so called "van der Meer" scan ($\sigma_{MB} = 71.4\pm7.1$ (syst.) mb).

In the forward region, in addition to the previous minimum-bias trigger conditions, the detection of at least one muon in the muon spectrometer is required. The muon trigger allows the selection of events where at least one particle, having $p_{\rm T}$ larger than the trigger threshold (0.5 GeV/c), has been detected in the muon trigger chambers \cite{ALICEatLHC}. 

The results presented in this report correspond to about 10$^8$ MB triggers (integrated luminosity of 1.5 nb$^{-1}$) and 10$^7$ muon triggers (11.6~nb$^{-1}$).

\section{Open charm measurement in hadronic channels}

The hadronic open charm decay signal extraction  is based on a invariant mass analysis of the fully reconstructed decay products. Among the most promising channels for open charm detection are the D$^0\rightarrow K^- \pi^+$ and D$^+\rightarrow K^- \pi^+\pi^+$ decays. 
In order to reduce the large combinatorial background, topological selections are applied. The selection of the events requires a displaced-vertex topology with a good alignment between the reconstructed D meson momentum and the flight path defined as the line between the primary and the secondary vertices. This selection is made possible thanks to the excellent tracking precision provided by the Inner Tracking System. A more detailed description of the analysis procedure can be found in reference \cite{Ortona}.

\begin{figure}
\centering
		\begin{minipage}[t]{\textwidth} 
			\begin{center}
\includegraphics[width=0.45\columnwidth]{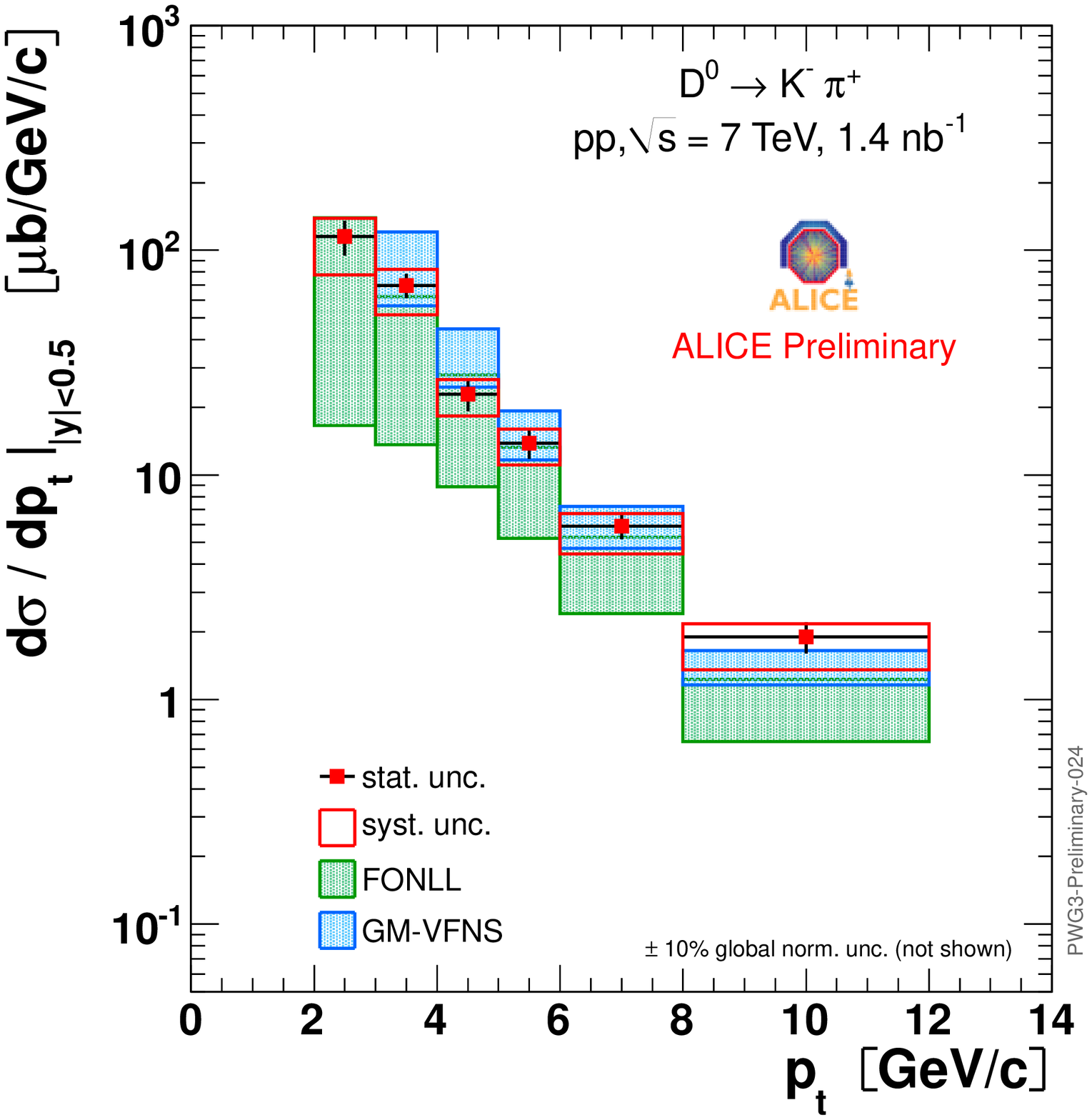}
				\hspace{0.8cm}
\includegraphics[width=0.45\columnwidth]{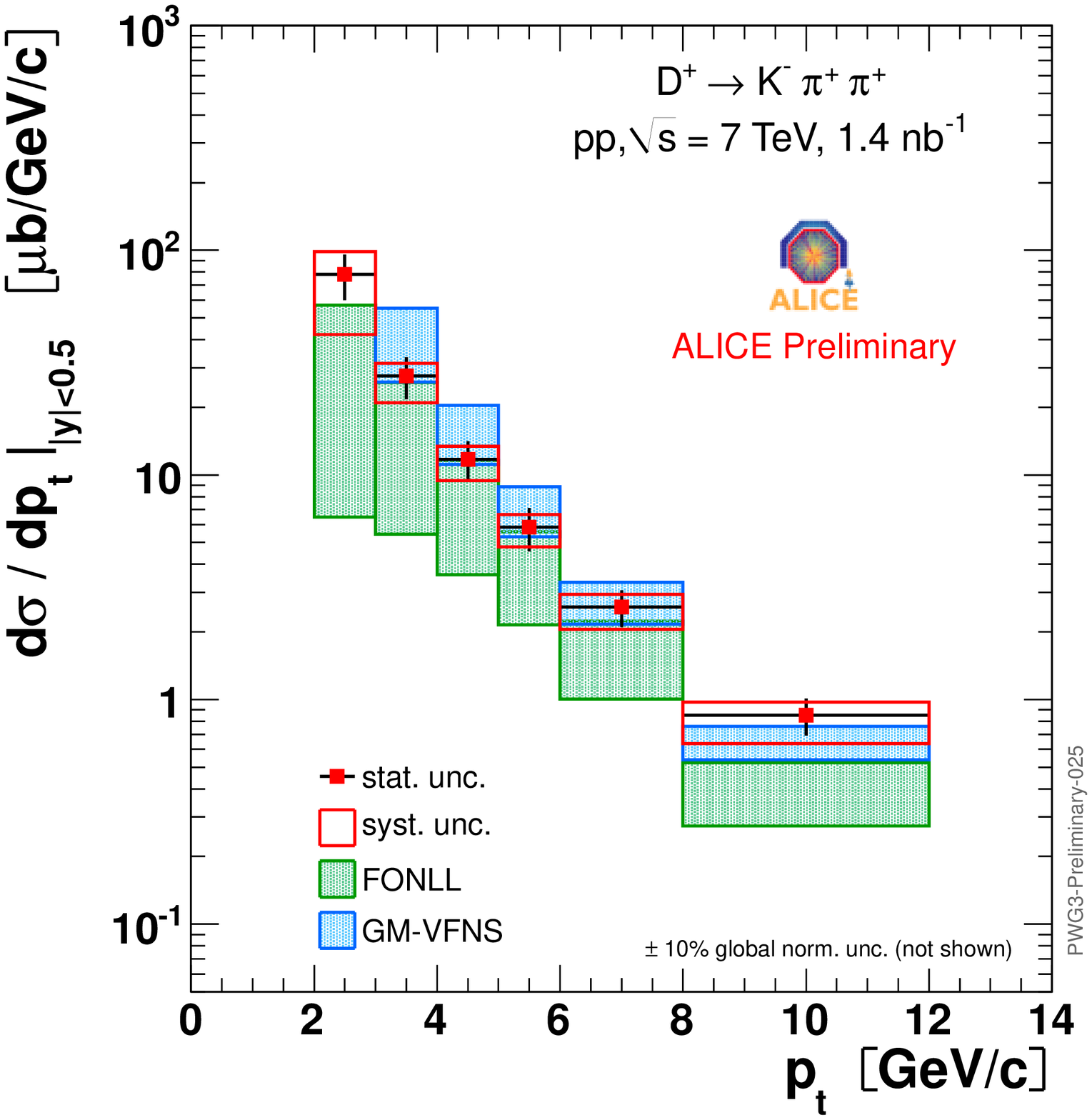}
			\end{center}
		\end{minipage}
\caption{$\frac{d\sigma}{dp_{\rm T}}$ cross sections for D$^0$ (left) and D$^+$ (right) mesons compared to pQCD calculations \cite{cacciari,Kniehl}.}
\label{fig:Dmeson}
\end{figure}

Using about 10$^8$ minimum-bias events, the $p_{\rm T}$-differential cross sections for D$^0$ and D$^+$ have been extracted with 6 $p_{\rm T}$ bins from 2 to 12~GeV/c. Figure~\ref{fig:Dmeson} shows the $\frac{d\sigma}{dp_{\rm T}}$ distributions measured in the rapidity range $|y|<0.5$ which is obtained by correcting the raw yields for efficiency, acceptance and feed down from beauty.  The correction for D mesons coming from B mesons feed-down has been computed using beauty over charm ratio given by the Fixed Order Next-to-Leading Log (FONLL) calculations \cite{cacciari}.

The correction for the beauty feed-down using
the measured displacement of the D mesons from beauty from the primary vertex is currently under way. The analysis is on going to exploit the entire 2010 data sample in order to extend the $p_{\rm T}$ coverage towards both larger and smaller $p_{\rm T}$ values.

\section{Electrons from heavy flavour decays}
Heavy flavour production at central rapidity can be studied using the single-electron $p_{\rm T}$ distribution \cite{Masciocchi}. The ALICE central detectors offer a very good particle identification capability. 
A clean electron sample is obtained applying the PID information given from TOF and TPC up to a momentum of 4 GeV/c. The remaining hadron contamination varies from the permille level at 0.5 GeV/c to about 15\% at 4 GeV/c and is subtracted by a multiple Gaussian fit performed in momentum slices of the TPC dE/dx distribution.
The signal extraction method consists in subtracting from the inclusive electron spectrum a cocktail of the background electron sources, namely, electrons from light hadron decays, photon conversions in the material of the beam pipe and of the SPD. After the cocktail subtraction, the inclusive cross section of electrons from charm and beauty decays is obtained. 

\begin{figure}
\centering
\includegraphics[width=0.7\columnwidth]{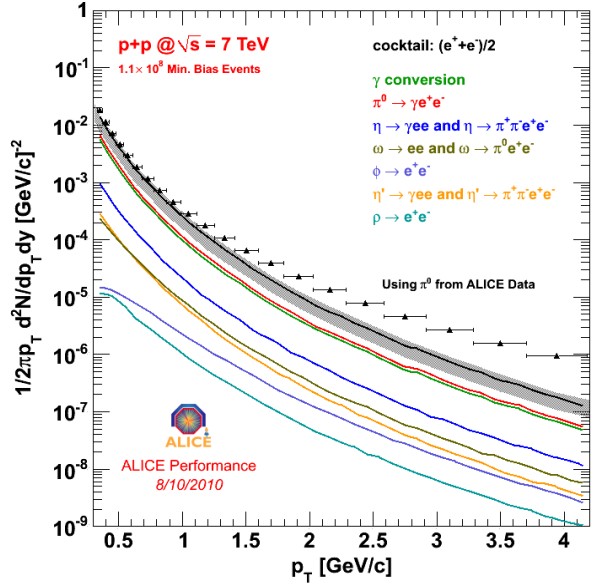}
\caption{Inclusive single electron yield as a function of $p_{\rm T}$ compared with the electron cocktail describing the main background sources.}
\label{fig:SingleElectron}
\end{figure}

Figure~\ref{fig:SingleElectron} shows the inclusive $p_{\rm T}$ electron spectrum in the transverse momentum range $0.4<p_{\rm T}<4$~GeV/c, after acceptance, efficiency and Bremsstrahlung effect corrections. Superimposed are the different cocktail contributions to this $p_{\rm T}$ distribution. The excess of electrons above the cocktail level can be attributed to the signal of electrons from heavy-flavour decays (including J/$\psi$, not yet included in the cocktail). An excess is clearly seen, increasing as a function of $p_{\rm T}$, as expected. The further developments of this analysis will include the electron PID from TRD and EMCal (Electromagnetic Calorimeter) to extend the analysis to higher $p_{\rm T}$ values.

\section{Forward single muons from heavy flavour decays}
\begin{figure}
\centering
\includegraphics[width=0.7\columnwidth]{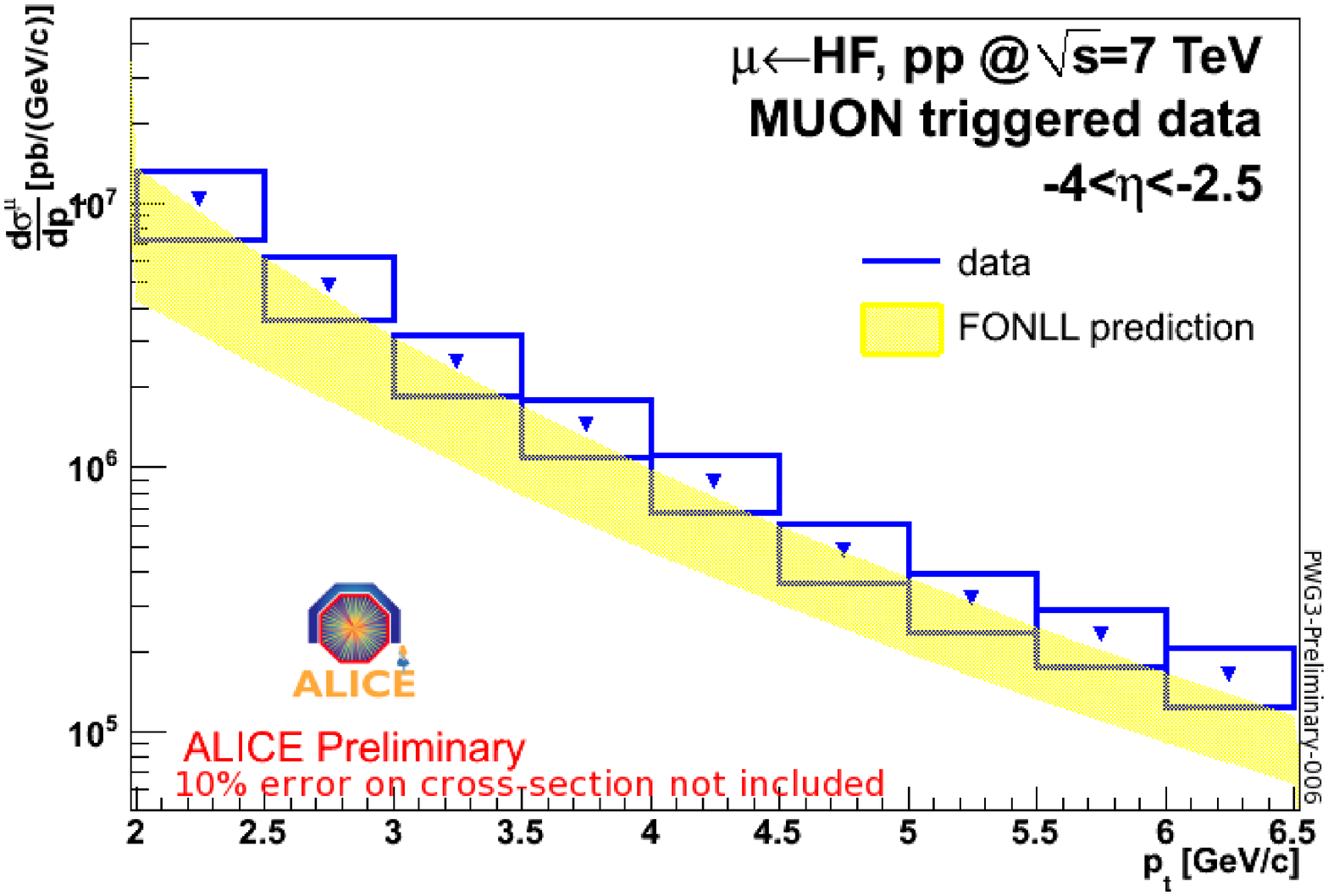}
\caption{Single muon coming from heavy flavour $\frac{d\sigma}{dp_{\rm T}}$ cross section compared to pQCD calculations.}
\label{fig:SingleMu}
\end{figure}

Heavy-flavour production at forward rapidity can be studied using the single-muon $p_{\rm T}$ distribution \cite{stocco}, measured in the muon spectrometer covering -4$<\eta<$-2.5. Apart from the heavy flavour, three main background sources are contributing to the single-muon $p_{\rm T}$ distribution. (i) Muons from the decay-in-flight of light hadrons produced at the interaction point called "decay muons". (ii) Muons from the decay of hadrons produced in the interaction with the front absorber, called "secondary muons". The last source of background are (iii) the punch-through hadrons. This last contribution can be rejected by requiring that the reconstructed track reaches the trigger stations placed behind the iron wall, leaving hits in at least three chambers out of four. In the present analysis, muons are required to have $p_{\rm T}$ greater than 2~GeV/c thus removing most of the background which has a softer $p_{\rm T}$ distributions. With such a cut, the contribution from secondary muons is reduced to about 3\%. About 25\% of the remaining background originates from decaying muons and is subtracted by means of simulations. The transverse momentum distribution of decay muons was generated with the Perugia-0 tune of PYTHIA \cite{Skands}, and normalized in such a way that the resulting fraction of decay-muons in the $p_{\rm T}$ range where this contribution is dominant (0.5$<p_{\rm T}<1$ GeV/c) is the same as the one provided by the simulation.
After background subtraction, the muon $p_{\rm T}$ spectrum is corrected for efficiency. The resulting differential cross section for charm and beauty production in the single muon decay channel is shown in figure~\ref{fig:SingleMu} for an integrated luminosity of 3.49 nb$^{-1}$. 
The comparison of a pQCD based theoretical calculation, shown in the figure,  shows a good agreement with data within the errors. 

\section{Inclusive J/$\psi$ production cross section}

Inclusive J/$\psi$ cross section can be measured in ALICE both at central and forward rapidity. At central rapidity, the J/$\psi$ are detected via the $e^+e^-$ decay channel in the rapidity range $|y|<0.88$. The particle identification is done using the TPC dE/dx information. The tracking is performed using the TPC and the ITS, requiring a minimum number of clusters inside the TPC of 90, in order to ensure a good quality track. The signal extraction is done subtracting the like sign distribution to the opposite sign invariant mass spectra. Preliminary results are based on an integrated luminosity of 1.5~nb$^{-1}$. The number of J/$\psi$  is extracted, with a bin
counting technique, in the mass range 2.9-3.15 GeV/c$^2$ and it amounts to $95\pm18 (stat.)$. After correction for acceptance and efficiency and normalization using the minimum-bias cross section, the resulting inclusive J/$\psi$ production cross section is: $d\sigma_{J/\psi}/dy_{(|y|<0.88)} = 7.36 \pm 1.22(stat.) \pm 1.32(syst.)^{+0.88}_{-1.84}(pol.)~\mu b$.  

At forward rapidity ($-4.0<y<-2.5$) the J/$\psi$ are measured via the $\mu^+\mu^-$ decay channel. The results presented in the following are based on an integrated luminosity of 11.6~nb$^{-1}$. The J/$\psi$ signal is extracted by fitting the  invariant mass distribution with two Crystal Ball functions (for the J/$\psi$ and the $\psi$') and two exponentials (below and above the J/$\psi$ mass).  A more detailed description of the analysis procedure can be found in reference~\cite{arnaldi}. The total number of J/$\psi$ is N$_{J/\psi}$ =1924 $\pm$77(stat.). The resulting inclusive J/$\psi$ production cross section is: $d\sigma_{J/\psi}/dy_{(-4<y|<-2.5)} = 7.25 \pm 0.29(stat.) \pm 0.98(syst.)^{+0.87}_{-1.50}(pol.)~\mu b$, after acceptance and efficiency corrections.
 
\begin{figure}
\centering
		\begin{minipage}[t]{\textwidth} 
			\begin{center}
\includegraphics[width=0.45\columnwidth]{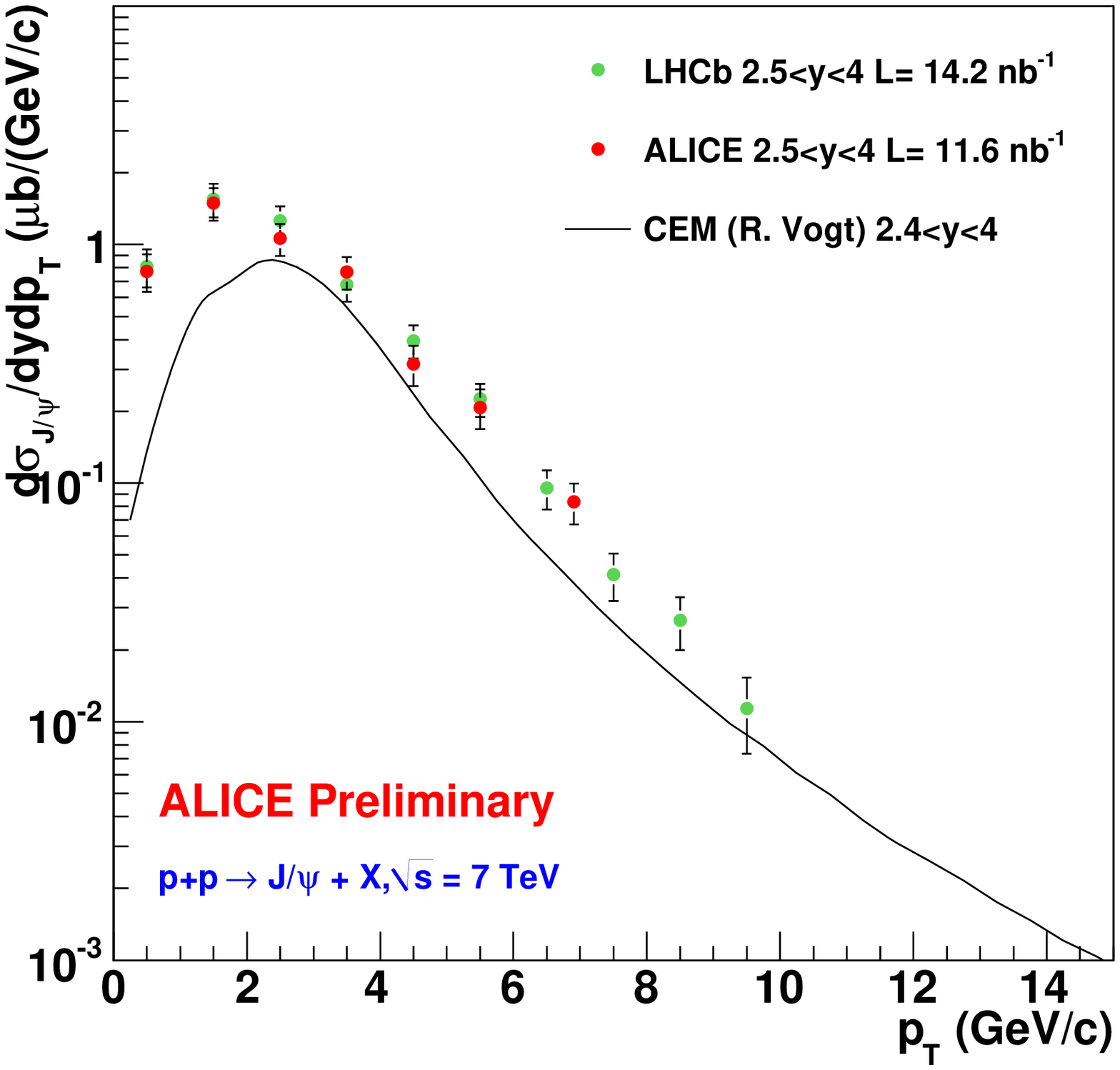}
				\hspace{0.8cm}
\includegraphics[width=0.45\columnwidth]{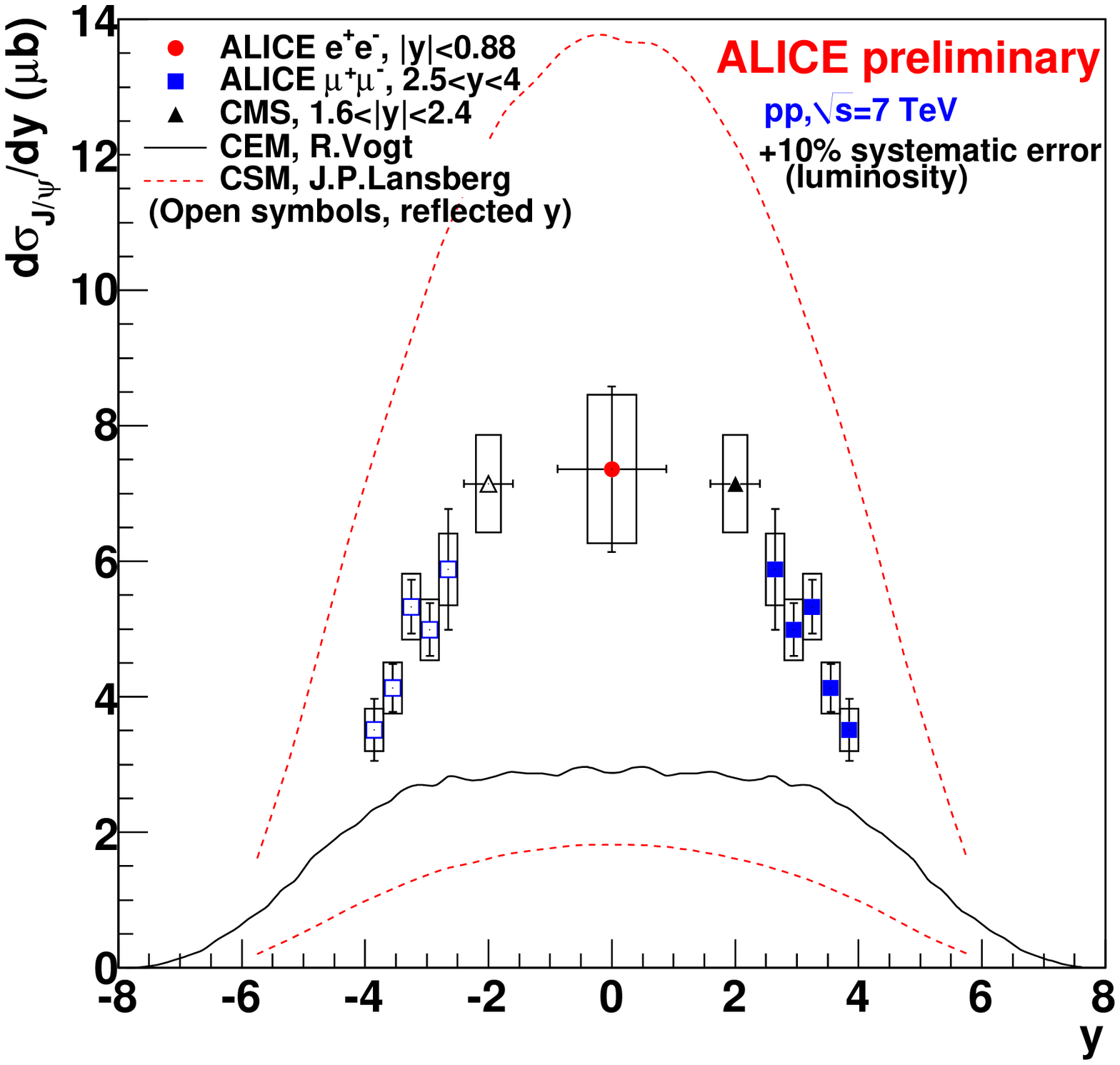}
			\end{center}
		\end{minipage}
\caption{Inclusive J/$\psi$ production cross sections as a function of $p_{\rm T}$ (left) in the range -4$<y<$-2.5 and as a function of rapidity (right). Comparison with results from other LHC experiments and pQCD calculations is also shown.  }
\label{fig:JPsi}
\end{figure}

The $p_{\rm T}$-differential cross section has been extracted using seven bins in $p_{\rm T}$ in the range 0$<p_{\rm T}<$8~GeV/c. Figure~\ref{fig:JPsi}-left shows the resulting $d\sigma_{J/\psi}/dp_{\rm T}$ in the rapidity range -4$<y<$-2.5. These results are compared with the corresponding measurement done by the LHCb experiment~\cite{LHCb}. The agreement with the preliminary results of the LHCb experiment, in the same rapidity range, is clearly visible. A comparison with a Color Evaporation Model (CEM) calculation for prompt J/$\psi$~\cite{vogt} is also shown. The predictions underestimate the experimental results but a more meaningful comparison would require the subtraction of J/$\psi$ coming from beauty decay.

The rapidity-differential cross section has been extracted using five rapidity bins and combined with the measurement done at central rapidity. Figure~\ref{fig:JPsi}-right shows the resulting $d\sigma_{J/\psi}/dy$ distribution. The values obtained in the forward region are also shown reflected with respect to y = 0. The ALICE measurement is compared to result from CMS~\cite{CMSJPsi}.  A comparison with the CEM calculation for prompt J/$\psi$  is shown as well as with a QCD-based Colour-Singlet Model~\cite{lansberg}. 

\section{Conclusions}

We have presented the first heavy-flavour production measurements performed by the ALICE experiment in proton-proton collisions at $\sqrt{s}$ = 7 TeV at LHC. ALICE has good capability to measure heavy-flavour in various decay channels at both central and forward rapidity down to zero $p_{\rm T}$. In particular, we have shown the first results concerning open heavy-flavour measurements in single electron and single muon channels as a function of $p_{\rm T}$. For both analysis, the next step will be to extend the $p_{\rm T}$ range. First comparison with pQCD calculations shows good agreement. We have measured charm production at central rapidity using fully-reconstructed decays of D mesons in hadronic final states. The D$^0$ and D$^+$ $p_{\rm T}$ distributions have been extracted. Analysis for other decay channels are well advanced. We have presented the first results on  rapidity- and $p_{\rm T}$-differential cross sections of inclusive J/$\psi$ production at central and forward rapidity, using its leptonic decay channels.

\end{document}